\documentstyle[twoside]{article}

\catcode`\@=11
\long\def\@makefntext#1{
\protect\noindent \hbox to 3.2pt {\hskip-.9pt  
$^{{\eightrm\@thefnmark}}$\hfil}#1\hfill}		

\def\@makefnmark{\hbox to 0pt{$^{\@thefnmark}$\hss}}	
	
\def\ps@myheadings{\let\@mkboth\@gobbletwo
\def\@oddhead{\hbox{}
\rightmark\hfil\eightrm\thepage}   
\def\@oddfoot{}\def\@evenhead{\eightrm\thepage\hfil
\leftmark\hbox{}}\def\@evenfoot{}
\def\sectionmark##1{}\def\subsectionmark##1{}}

\oddsidemargin=\evensidemargin
\newcounter{sectionc}\newcounter{subsectionc}\newcounter{subsubsectionc}
\renewcommand{\section}[1] {\vspace{12pt}\addtocounter{sectionc}{1} 
\setcounter{subsectionc}{0}\setcounter{subsubsectionc}{0}\noindent 
	{\tenbf\thesectionc. #1}\par\vspace{5pt}}
\renewcommand{\subsection}[1] {\vspace{12pt}\addtocounter{subsectionc}{1} 
	\setcounter{subsubsectionc}{0}\noindent 
	{\bf\thesectionc.\thesubsectionc. {\kern1pt \bfit #1}}\par\vspace{5pt}}
\renewcommand{\subsubsection}[1] {\vspace{12pt}\addtocounter{subsubsectionc}{1}
	\noindent{\tenrm\thesectionc.\thesubsectionc.\thesubsubsectionc.
	{\kern1pt \tenit #1}}\par\vspace{5pt}}

\newcounter{appendixc}
\newcounter{subappendixc}[appendixc]
\newcounter{subsubappendixc}[subappendixc]
\renewcommand{\thesubappendixc}{\Alph{appendixc}.\arabic{subappendixc}}
\renewcommand{\thesubsubappendixc}
	{\Alph{appendixc}.\arabic{subappendixc}.\arabic{subsubappendixc}}

\renewcommand{\appendix}[1] {\vspace{12pt}
        \refstepcounter{appendixc}
        \setcounter{figure}{0}
        \setcounter{table}{0}
        \setcounter{lemma}{0}
        \setcounter{theorem}{0}
        \setcounter{corollary}{0}
        \setcounter{definition}{0}
        \setcounter{equation}{0}
        \renewcommand{\thefigure}{\Alph{appendixc}.\arabic{figure}}
        \renewcommand{\thetable}{\Alph{appendixc}.\arabic{table}}
        \renewcommand{\theappendixc}{\Alph{appendixc}}
        \renewcommand{\thelemma}{\Alph{appendixc}.\arabic{lemma}}
        \renewcommand{\thetheorem}{\Alph{appendixc}.\arabic{theorem}}
        \renewcommand{\thedefinition}{\Alph{appendixc}.\arabic{definition}}
        \renewcommand{\thecorollary}{\Alph{appendixc}.\arabic{corollary}}
        \renewcommand{\theequation}{\Alph{appendixc}.\arabic{equation}}
        \noindent{\tenbf Appendix \theappendixc #1}\par\vspace{5pt}}
\newcommand{\subappendix}[1] {\vspace{12pt}
        \refstepcounter{subappendixc}
        \noindent{\bf Appendix \thesubappendixc. {\kern1pt \bfit #1}}
	\par\vspace{5pt}}
\newcommand{\subsubappendix}[1] {\vspace{12pt}
        \refstepcounter{subsubappendixc}
        \noindent{\rm Appendix \thesubsubappendixc. {\kern1pt \tenit #1}}
	\par\vspace{5pt}}

\newcommand{\textlineskip}{\baselineskip=13pt}
\newcommand{\smalllineskip}{\baselineskip=10pt}

\def\eightcirc{
\begin{picture}(0,0)
\put(4.4,1.8){\circle{6.5}}
\end{picture}}
\def\eightcopyright{\eightcirc\kern2.7pt\hbox{\eightrm c}} 
\newcommand{\pub}[1]{{\begin{center}\footnotesize\smalllineskip 
	\\
	\end{center}
	}}

\def\abstracts#1#2#3{{
	\centering{\begin{minipage}{4.5in}\baselineskip=10pt\footnotesize
	\parindent=0pt #1\par 
	\parindent=15pt #2\par
	\parindent=15pt #3
	\end{minipage}}\par}} 

\newcommand{\bibit}{\nineit}

\renewenvironment{thebibliography}[1]
	{\frenchspacing
	 \ninerm\baselineskip=11pt
	 \begin{list}{\arabic{enumi}.}
        {\usecounter{enumi}\setlength{\parsep}{0pt}     
	 \setlength{\leftmargin 12.7pt}{\rightmargin 0pt} 
         \setlength{\itemsep}{0pt} \settowidth
	{\labelwidth}{#1.}\sloppy}}{\end{list}}

\newcounter{itemlistc}
\newcounter{romanlistc}
\newcounter{alphlistc}
\newcounter{arabiclistc}

\newcommand{\fcaption}[1]{
        \refstepcounter{figure}
        \setbox\@tempboxa = \hbox{\footnotesize Fig.~\thefigure. #1}
        \ifdim \wd\@tempboxa > 5in
           {\begin{center}
        \parbox{5in}{\footnotesize\smalllineskip Fig.~\thefigure. #1}
            \end{center}}
        \else
             {\begin{center}
             {\footnotesize Fig.~\thefigure. #1}
              \end{center}}
        \fi}

\newcommand{\tcaption}[1]{
        \refstepcounter{table}
        \setbox\@tempboxa = \hbox{\footnotesize Table~\thetable. #1}
        \ifdim \wd\@tempboxa > 5in
           {\begin{center}
        \parbox{5in}{\footnotesize\smalllineskip Table~\thetable. #1}
            \end{center}}
        \else
             {\begin{center}
             {\footnotesize Table~\thetable. #1}
              \end{center}}
        \fi}

\def\@citex[#1]#2{\if@filesw\immediate\write\@auxout
	{\string\citation{#2}}\fi
\def\@citea{}\@cite{\@for\@citeb:=#2\do
	{\@citea\def\@citea{,}\@ifundefined
	{b@\@citeb}{{\bf ?}\@warning
	{Citation `\@citeb' on page \thepage \space undefined}}
	{\csname b@\@citeb\endcsname}}}{#1}}

\newif\if@cghi
\def\cite{\@cghitrue\@ifnextchar [{\@tempswatrue
	\@citex}{\@tempswafalse\@citex[]}}
\def\citelow{\@cghifalse\@ifnextchar [{\@tempswatrue
	\@citex}{\@tempswafalse\@citex[]}}
\def\@cite#1#2{{$\null^{#1}$\if@tempswa\typeout
	{IJCGA warning: optional citation argument 
	ignored: `#2'} \fi}}

\def\pmb#1{\setbox0=\hbox{#1}
	\kern-.025em\copy0\kern-\wd0
	\kern.05em\copy0\kern-\wd0
	\kern-.025em\raise.0433em\box0}

\def\fnm#1{$^{\mbox{\scriptsize #1}}$}
\def\fnt#1#2{\footnotetext{\kern-.3em
	{$^{\mbox{\scriptsize #1}}$}{#2}}}

\def\fpage#1{\begingroup
\voffset=.3in
\thispagestyle{empty}\begin{table}[b]\centerline{\footnotesize #1}
	\end{table}\endgroup}

\def\runninghead#1#2{\pagestyle{myheadings}
\markboth{{\protect\footnotesize\it{\quad #1}}\hfill}
{\hfill{\protect\footnotesize\it{#2\quad}}}}
\font\tenrm=cmr10
\font\tenit=cmti10 
\font\tenbf=cmbx10
\font\bfit=cmbxti10 at 10pt
\font\ninerm=cmr9
\font\nineit=cmti9

\font\eightrm=cmr8

\def\qed{\hbox{${\vcenter{\vbox{			
   \hrule height 0.4pt\hbox{\vrule width 0.4pt height 6pt
   \kern5pt\vrule width 0.4pt}\hrule height 0.4pt}}}$}}

\begin{document}

\runninghead{Search for Anomaly at High x $\ldots$} 
{Search for Anomaly at High x $\ldots$}

\normalsize\textlineskip
\thispagestyle{empty}
\setcounter{page}{1}

\vspace*{0.88truein}

\fpage{1}
\centerline{\bf SEARCH FOR ANOMALY AT HIGH X IN POLARIZED}
\vspace*{0.035truein}
\centerline{\bf DEEP INELASTIC SCATTERING DATA}
\vspace*{0.37truein}
\centerline{\footnotesize WOJCIECH WI\'SLICKI\fnm{*}\fnt{*}{email: wislicki@fuw.edu.pl}}
\vspace*{0.015truein}
\centerline{\footnotesize\it So\l tan Institute for Nuclear Studies}
\baselineskip=10pt
\centerline{\footnotesize\it Ho\.za 69, PL-00-681 Warsaw}
\vspace*{10pt}
\vspace*{0.225truein}

\vspace*{0.21truein}
\abstracts{
An idea of possible anomalous contribution of non-perturbative origin to the nucleon 
spin was examined by analysing data on spin asymmetries in polarized deep inelastic 
scattering of leptons on nucleons. The region of high Bjorken $x$ was explored.
It was shown that experimental data available at present do not exhibit any evidence
for this effect.}{}{}

\vspace*{10pt}

\textlineskip			
\vspace*{12pt}			

As one of possible explanations of the nucleon spin problem it was suggested \cite{altarelli_1,efremov_1}
that the overall quark spin in the nucleon may be diminished by the axial anomaly term.
If the anomaly is related to the gluon polarization then it is observable at small $x$ only.
However, as discussed in refs.\cite{bass1,bass2}, there is no {\it a priori} argument that the anomaly must be the small-$x$
effect only. 
For the region of high $x$ the model of the nucleon as consisting of three massive constituent quarks is well justified. 
The valence quarks move in a non-perturbative colour background field which could be identified with the anomaly and possess a non-trivial spin structure. 
The problem of effective spin dilution by a colour field in the constituent-like picture of the nucleon is sometimes addressed in the context of non-perturbative models of hadrons and is worth of being investigated experimentally.

In order to estimate the effect of a possible colour screening of the quarks spin in the nucleon we analyse existing data on the inclusive and semi-inclusive asymmetries measured in polarized deep inelastic scattering experiments. 
A direct way to do this would be a comparison of the quark spin distributions obtained from the C-even spin observables, e.g. $g_1$ or single-charge semi-inclusive asymmetries, where the anomaly contributes, with the anomaly free C-odd quantities, as $g_3$ or the semi-inclusive asymmetries for difference of charge \cite{bass1}. 
Any inconsistency between them would eventually evidence for a large-$x$ anomaly.
However, the only data available at present are C-even \cite{emc1}$^{-}$\cite{e154}. 
Therefore the method we use here is to express C-even asymmetries by quark spin distributions, using the parametrization of gauge-invariant quark spins, modified by the anomaly term as proposed in refs. \cite{bass1,bass2}, and fit them to all existing data.

It was shown that in order to make the quark spin distributions gauge independent, and thus ensure conservation of chirality, one has to modify the quark helicity distributions $q(x,Q^2)^{\uparrow}$ and $q(x,Q^2)^{\downarrow}$ 
\begin{eqnarray}
\tilde{q}(x,Q^2)^{\uparrow} & = & q(x,Q^2)^{\uparrow}+\frac{1}{4}\kappa(x,Q^2) \nonumber \\
\tilde{q}(x,Q^2)^{\downarrow} & = & q(x,Q^2)^{\downarrow}-\frac{1}{4}\kappa(x,Q^2) 
\label{eq01}
\end{eqnarray}
for both quarks and antiquarks and for each quark flavour. 
In the following all quark distributions will refer to the proton.

In experiments on polarized deep inelastic scattering of leptons on nucleons one determines the cross section asymmetries \cite{emc1}
\begin{equation}
A(x,Q^2)=\frac{\sigma_a(x,Q^2)-\sigma_p(x,Q^2)}{\sigma_a(x,Q^2)+\sigma_p(x,Q^2)}.
\label{eq02}
\end{equation}
The indices $_a$ and $_p$ refer to the sum of spins of the virtual photon and the nucleon for two opposite polarizations of the target nucleon.
For the proton and neutron target $a=1/2$ and $p=3/2$. In case of deuteron target these spins are equal to $0$ and $2$, respectively.\fnm{a}\fnt{a}{The deuteron cross section is considered to be the sum of the proton and the neutron cross sections, corrected for the D-state of the deuteron as in ref. \cite{smc3}} 
In the inclusive measurement the cross sections in (\ref{eq02}) are the inclusive polarized deep inelastic cross sections and for the semi-inclusive experiments they refer to polarized cross sections for leptoproduction of hadrons in the current fragmentation region.

In the region of high $x$ the non-strange sea, the strange sea and the ratio of the transverse to the longitudinal photoabsorption cross sections are negligible. 
Assuming anomaly (\ref{eq01}) the inclusive asymmetries for the proton, the neutron and the deuteron targets are given in the quark-parton model (QPM) as
\begin{equation}
A_1^p(x,Q^2)=\frac{\frac{2}{9}\Delta u(x,Q^2)+\frac{1}{18}\Delta d(x,Q^2)+\frac{1}{3}\kappa(x,Q^2)}
                  {\frac{2}{9}u(x,Q^2)+\frac{1}{18}d(x,Q^2)},
\label{eq03}
\end{equation}
\begin{equation}
A_1^n(x,Q^2)=\frac{\frac{1}{18}\Delta u(x,Q^2)+\frac{2}{9}\Delta d(x,Q^2)+\frac{1}{3}\kappa(x,Q^2)}
                  {\frac{1}{18}u(x,Q^2)+\frac{2}{9}d(x,Q^2)}
\label{eq04}
\end{equation}
and
\begin{equation}
A_1^d(x,Q^2)=\frac{\frac{5}{18}[\Delta u(x,Q^2)+\Delta d(x,Q^2)]+\frac{2}{3}\kappa(x,Q^2)}
                  {\frac{5}{18}[u(x,Q^2)+d(x,Q^2)]}.
\label{eq05}
\end{equation}
The QPM analysis of semi-inclusive asymmetries requires knowledge about fragmentation functions of quarks into hadrons $D_q^h(z,Q^2)$. 
Since not all of them are measured one has to reduce the number of independent fragmentation functions by using charge conjugation and isospin symmetry and some additional assumptions for favoured and unfavoured fragmentation functions
\cite{smc1,emc2}
\begin{eqnarray}
D_1 & \equiv & D_u^{\pi^+}=D_{\bar d}^{\pi^+}=D_{\bar u}^{\pi^-}=D_d^{\pi^-}=D_s^{K^-}=D_{\bar s}^{K^+} \nonumber \\
D_2 & \equiv & D_u^{\pi^-}=D_{\bar d}^{\pi^-}=D_{\bar u}^{\pi^+}=D_d^{\pi^+}=D_s^{\pi^+}=D_{\bar s}^{\pi^+}=D_s^{\pi^-}=D_{\bar s}^{\pi^-} \nonumber \\
D_3 & \equiv & D_u^{K^+}=D_{\bar u}^{K^-} \nonumber \\
D_4 & \equiv & D_u^{K^-}=D_{\bar d}^{K^-}=D_d^{K^-}=D_{\bar s}^{K^-}=D_{\bar u}^{K^+}=D_d^{K^+}=D_{\bar d}^{K^+}=D_s^{K^+} \nonumber \\
D_5 & \equiv & D_u^p=D_d^p=D_{\bar u}^{\bar p}=D_{\bar d}^{\bar p} \nonumber \\
D_6 & \equiv & D_u^{\bar p}=D_d^{\bar p}=D_s^{\bar p}=D_{\bar s}^{\bar p}=D_{\bar u}^{p}=D_{\bar d}^{p}=D_s^{p}=D_{\bar s}^{p}. 
\label{eq06}
\end{eqnarray}
We assume that fragmentation functions do not depend on quark helicity and that
the fragmentation of a parton factorize from its distribution. 
The factorization is well confirmed for quarks at high $Q^2$ and in the current fragmentation region which is effectively selected by cuting on the Feynman variable, $x_F>0$, or on the fraction of the virtual photon energy taken by the final state hadron, $z=E_h/\nu \gg 0$.
Factorization was never studied for the anomaly and it remains an assumption additional to the QPM.
Semi-inclusive data of EMC \cite{emc1} and SMC \cite{smc1,smc4} were taken without hadron identification and only the electric charge of a hadron was known. 
Therefore the semi-inclusive asymmetries have to be derived for the mixture of pions, kaons and protons in the final state.
Spin asymmetries for the production of positive (negative) hadrons on the proton, neutron and deuteron at high $x$ are the following:
\begin{eqnarray}
A_{1\,p}^{+(-)}(x,Q^2) & = & \frac{\frac{4}{9}D_u^{+(-)}\Delta u(x,Q^2)+\frac{1}{9}D_d^{+(-)}\Delta d(x,Q^2)+\frac{2}{3}D_{\kappa}\kappa(x,Q^2)}
                        {\frac{4}{9}D_u^{+(-)}u(x,Q^2)+\frac{1}{9}D_d^{+(-)}d(x,Q^2)}, 
\label{eq07}
\end{eqnarray}
\begin{eqnarray}
A_{1\,n}^{+(-)}(x,Q^2) & = & \frac{\frac{1}{9}D_d^{+(-)}\Delta u(x,Q^2)+\frac{4}{9}D_u^{+(-)}\Delta d(x,Q^2)+\frac{2}{3}D_{\kappa}\kappa(x,Q^2)}
                        {\frac{1}{9}D_d^{+(-)}u(x,Q^2)+\frac{4}{9}D_u^{+(-)}d(x,Q^2)}
\label{eq08}
\end{eqnarray}
and
\begin{eqnarray}
A_{1\,d}^{+(-)}(x,Q^2) & = & \frac{\frac{1}{9}[4D_u^{+(-)}+D_d^{+(-)}][\Delta u(x,Q^2)+\Delta d(x,Q^2)]+\frac{4}{3}D_{\kappa}\kappa(x,Q^2)}
                        {\frac{1}{9}[4D_u^{+(-)}+D_d^{+(-)}][u(x,Q^2)+d(x,Q^2)]},
\label{eq09}
\end{eqnarray}
where
\begin{eqnarray}
D_u^+ & \equiv & D_1+D_3+D_5 \nonumber \\
D_d^+ & \equiv & D_2+D_4+D_5 \nonumber \\
D_u^- & \equiv & D_2+D_4+D_6 \nonumber \\
D_d^- & \equiv & D_1+D_4+D_6. 
\label{eq10}
\end{eqnarray}
In eqs. (\ref{eq07}-\ref{eq10}) $D_i\equiv \int_{z_0}^1\,dz\,D_i(z,Q^2),\;\;\;{\small i=1,\ldots,6}$ ($z_0=0.2$ in ref. \cite{smc4} and $0.1$ in ref. \cite{emc1}), and $D_{\kappa}$ is the fragmentation function of the anomaly.\fnm{b}\fnt{b}{The
fragmentation function of the anomaly can be different than the fragmentation function of
a quark into the same hadron. The assumption of their equality, in our opinion unjustified, was implicitly done in ref. \cite{gullenstern1}}
It is expected to be the same for the positive and negative hadrons and is defined as\fnm{c}\fnt{c}{In ref. \cite{bass1} the anomaly fragmentation function refers to any charged hadron and not to the specific charge as in our case}
\begin{eqnarray}
D_{\kappa} & \equiv & D_{\kappa}^{\pi^+}+D_{\kappa}^{K^+}+D_{\kappa}^{p} \nonumber \\
           &    =   & D_{\kappa}^{\pi^-}+D_{\kappa}^{K^-}+D_{\kappa}^{\bar p}. 
\label{eq11}
\end{eqnarray}
The fragmentation function of the anomaly $D_{\kappa}$ can be determined, as suggested in ref.~\cite{bass1}, by comparing the semi-inclusive asymmetry for all hadrons, including positive, negative and neutral\fnm{d}\fnt{d}{Taking only positive
and negative hadrons, as in ref.~\cite{bass1}, is not sufficient.}, on the deuteron, $A_{1\,d}^{sum}$, to the inclusive asymmetry $A_1^d(x,Q^2)=\frac{g_1^d(x,Q^2)}{F_1^d(x,Q^2)}$, 
which are expected to be equal. 
To account for neutral particles, for which fragmentation functions are poorly known, we use again the isospin symmetry and charge conjugation to relate them to measured functions $D_{1,\ldots,6}$. 
For neutral hadrons not heavier than the nucleon those relations are \cite{andersson_1}:
\begin{eqnarray}
D_u^{\pi^0} & = & D_{\bar u}^{\pi^0}=D_d^{\pi^0}=D_{\bar d}^{\pi^0}=\frac{1}{2}(D_1+D_2) \nonumber \\
D_s^{\pi^0} & = & D_{\bar s}^{\pi^0}=D_2 \nonumber \\
D_u^{\eta} & = & D_{\bar u}^{\eta}=D_d^{\eta}=D_{\bar d}^{\eta}=\frac{1}{6}(D_1+2\,D_2) \nonumber \\
D_s^{\eta} & = & D_{\bar s}^{\eta}=\frac{1}{3}(D_2+D_3) \nonumber \\
D_u^{\omega} & = & D_{\bar u}^{\omega}=D_d^{\omega}=D_{\bar d}^{\omega}=\frac{3}{2}(D_1+D_2) \nonumber \\
D_s^{\omega} & = & D_{\bar s}^{\omega}=3\,D_2 \nonumber \\
D_u^{K_S} & = & D_{\bar u}^{K_S}=D_u^{K_L}=D_{\bar u}^{K_L}=D_4 \nonumber \\
D_d^{K_S} & = & D_{\bar d}^{K_S}=D_d^{K_L}=D_{\bar d}^{K_L}=\frac{1}{2}(D_3+D_4) \nonumber \\
D_s^{K_S} & = & D_{\bar s}^{K_S}=D_s^{K_L}=D_{\bar s}^{K_L}=\frac{1}{2}(D_1+D_4) \nonumber \\
D_u^n & = & D_d^n=D_{\bar u}^{\bar n}=D_{\bar d}^{\bar n}=D_5 \nonumber \\
D_{\bar u}^n & = & D_{\bar d}^n=D_s^n=D_{\bar s}^n=D_u^{\bar n}=D_d^{\bar n}=D_s^{\bar n}=D_{\bar s}^{\bar n}=D_6.
\label{eq12}
\end{eqnarray}
The asymmetry $A_{1\,d}^{sum}$ is equal to
\begin{equation}
A_{1\,d}^{sum}(x,Q^2)=\frac{\frac{1}{9}{\cal A}[\Delta u(x,Q^2)+\Delta d(x,Q^2)]+\frac{4}{3}{\cal B}\kappa(x,Q^2)}
                      {\frac{1}{9}{\cal A}[u(x,Q^2)+d(x,Q^2)]},
\label{eq13}
\end{equation}
where the coefficients $\cal A$ and $\cal B$ depend on the fragmentation functions (\ref{eq12}) and are equal to
\begin{eqnarray}
{\cal A} & = & \frac{80}{3}D_1+\frac{85}{3}D_2+9D_3+\frac{37}{2}D_4+18D_5+12D_6 \nonumber \\
{\cal B} & \simeq & 1.68\,D_{\kappa}.
\label{eq14}
\end{eqnarray}
The asymmetry $A_{1\,d}^{sum}$ (\ref{eq13}) can be rewritten in the form
\begin{equation}
A_{1\,d}^{sum}(x,Q^2)=\frac{g_1^d(x,Q^2)}{F_1^d(x,Q^2)}+\frac{\frac{2}{3}\kappa(x,Q^2)}{F_1^d(x,Q^2)}
                      \left[\frac{5{\cal B}}{{\cal A}}-1\right],
\label{eq17}
\end{equation}
hence
\begin{equation}
D_{\kappa}\simeq 0.12\,{\cal A}.
\label{eq18}
\end{equation}

In our analysis we used the inclusive asymmetries on proton from EMC \cite{emc1}, SMC~\cite{smc2} and E143 \cite{e143_1}, on deuteron from SMC \cite{smc3} and E143 \cite{e143_2} and on neutron from E142 \cite{e142}, HERMES \cite{hermes_1} and E154 \cite{e154} experiments. 
We used also single-charge semi-inclusive asymmetries on proton from EMC \cite{emc1} and on proton and deuteron from SMC \cite{smc4}. 
We considered only range of $x>0.15$.
If more than one data set is available for given asymmetry the data were combined in SMC bins.
For each bin of $x$ eqs.~(\ref{eq03}),(\ref{eq04}),(\ref{eq05}),(\ref{eq07}) and (\ref{eq09}) constitute a system of 7 linear equations for 3 spin distributions $\Delta u$, $\Delta d$ and $\kappa$. 
Asymmetries were assumed to be independent of $Q^2$, consistent with ithe SMC and E143 observations for the inclusive case \cite{smc2,smc3,e143_2}. 
Any possible effects of scale mixing by anomaly were neglected. 
For quark distribution functions the GRV parametrization \cite{grv1} at $Q^2=10$ GeV$^2$ was used and for fragmentation functions we used EMC data \cite{emc2}. 
Three unknown spin distributions, $\Delta u(x)$, $\Delta d(x)$ and $\kappa(x)$, were evaluated from the system of seven equations by the least squares method, using the full correlation matrix between asymmetries \cite{smc4}\fnm{e}\fnt{e}{For EMC data \cite{emc1} 
the covariance matrix is not known. It was assumed that correlations are the same as for SMC proton data and the influence of this assumption on values of polarized distributions was found to be negligible.} 
The distributions of $\Delta u(x)$, $\Delta d(x)$ and $\kappa(x)$ are given in tab.~1\fnm{f}\fnt{f}{In evaluation of systematic errors the correlations between systematic errors of asymmetries were unknown. The correlation coefficients were varied between 0\% and 100\% and it was found that the maximum change of the systematic error of the anomaly was 12\%. 
This contribution was added to the systematic error.}.
\begin{table}[h]
\tcaption{The values of $\Delta u$, $\Delta d$ and $\kappa$ obtained at $<x>$ for $x>0.15$. The first error is statistical and the second is systematic}
\centerline{\footnotesize\smalllineskip
\begin{tabular}{|c|c|c|c|c|} \hline
$x_L-x_U$ & $<x>$ & $\Delta u(x)$ & $\Delta d(x)$ & $\kappa(x)$ \\ \hline
$0.15-0.2$ & $0.17$ & $1.159 \pm 0.092 \pm 0.037$ & $-0.636 \pm 0.084 \pm 0.030$ & $0.051 \pm 0.058 \pm 0.011$ \\
$0.2-0.3$  & $0.24$ & $0.919 \pm 0.067 \pm 0.027$ & $-0.535 \pm 0.065 \pm 0.024$ & $0.065 \pm 0.045 \pm 0.009$ \\
$0.3-0.4$  & $0.34$ & $0.693 \pm 0.076 \pm 0.022$ & $-0.185 \pm 0.073 \pm 0.020$ & $-0.009 \pm 0.050 \pm 0.004$ \\
$0.4-0.7$  & $0.48$ & $0.376 \pm 0.045 \pm 0.011$ & $-0.156 \pm 0.047 \pm 0.007$ & $0.021 \pm 0.031 \pm 0.004$ \\ \hline
\end{tabular}}
\end{table}
The values of integrals of $\Delta u(x)$, $\Delta d(x)$ and $\kappa(x)$ are listed in tab.~2. 
Contributions from the unmeasured region $x>0.7$ were estimated from the extrapolation of the function $Ax^B(1-x)^C$ determined from the fit to measured points at $0.15<x<0.7$.
We observe that the anomaly distribution $\kappa(x)$ does not exhibit any significant
deviation from zero. 
The integral $\int_{0.15}^{1}dx\,\kappa(x)$ is consistent with null hypothesis within one standard deviation.

In addition to ongoing experiments at DESY (HERMES) and at SLAC (E155), which will provide new data on both inclusive and hadron asymmetries, substantial reduction of the statistical error in the region od high $x$ is expected in the COMPASS experiment, presently under construction at CERN \cite{compass}. 
Particles scattered and produced on the proton and deuteron polarized targets at angles up to 180 mrad will be detected by a dedicated spectrometer equiped with a magnet of wide aperture and thus enriching event statistics at high $x$. 
Adding to the present analysis the inclusive and semi-inclusive asymmetries measured with the precission foreseen at high $x$ in COMPASS one gets the statistical error of the anomaly term $\kappa$ equal to $\pm 0.004$.
\begin{table}[h]
\tcaption{Integrals of $\Delta u$, $\Delta d$ and $\kappa$}
\centerline{\footnotesize\smalllineskip
\begin{tabular}{|c|c|c|c|} \hline
$x$                    & $0.15-0.7$ & $0.7-1$ & 0.15-1 \\ \hline
$\int\,dx\,\Delta u(x)$ & $0.396 \pm 0.018 \pm 0.005$ & $0.048 \pm 0.002 \pm 0.001$ & $0.444 \pm 0.019 \pm 0.005$ \\
$\int\,dx\,\Delta d(x)$ & $-0.184 \pm 0.018 \pm 0.004$ & $-0.024 \pm 0.002 \pm 0.001$ & $-0.208 \pm 0.018 \pm 0.004$\\
$\int\,dx\,\kappa(x)$  & $0.013 \pm 0.012 \pm 0.002$ & $0.002 \pm 0.002 \pm (3 \times 10^{-4})$ & $0.015 \pm 0.013 \pm 0.002$ \\ \hline
\end{tabular}}
\end{table}

\end{document}